\documentclass[aip,reprint]{revtex4-2}

\usepackage{graphicx}% Include figure files
\usepackage{dcolumn}% Align table columns on decimal point
\usepackage{bm}% bold math
\usepackage[utf8]{inputenc}
\usepackage[T1]{fontenc}
\usepackage{mathptmx}
\usepackage{etoolbox}
\usepackage[colorlinks,linkcolor=blue,anchorcolor=red,citecolor=blue,urlcolor=blue]{hyperref}
\draft % marks overfull lines with a black rule on the right

\makeatletter
\def\@email#1#2{%
 \endgroup
 \patchcmd{\titleblock@produce}
  {\frontmatter@RRAPformat}
  {\frontmatter@RRAPformat{\produce@RRAP{*#1\href{mailto:#2}{#2}}}\frontmatter@RRAPformat}
  {}{}
}%
\makeatother

\begin{document}

\title{Time-, spin-, and angle-resolved photoemission spectroscopy \\with a 1-MHz 10.7-eV pulse laser} %Title of pape

\author{Kaishu~Kawaguchi}
\altaffiliation{These authors contributed equally.}
\email{kuroken224@hiroshima-u.ac.jp}
\affiliation{Institute for Solid State Physics (ISSP), The University of Tokyo, Kashiwa, Chiba 277-8581, Japan}

\author{Kenta~Kuroda*}
\altaffiliation{These authors contributed equally.}
\affiliation{Institute for Solid State Physics (ISSP), The University of Tokyo, Kashiwa, Chiba 277-8581, Japan}
\affiliation{Graduate School of Advanced Science and Engineering, Hiroshima University, Higashi-hiroshima, Hiroshima 739-8526, Japan}
\affiliation{International Institute for Sustainability with Knotted Chiral Meta Matter (WPI-SKCM${}^{2}$), Hiroshima University, Higashi-hiroshima, Hiroshima 739-8526, Japan}

\author{Z.~Zhao}
\affiliation{Institute for Solid State Physics (ISSP), The University of Tokyo, Kashiwa, Chiba 277-8581, Japan}
\affiliation{School of Information Science and Engineering, Shandong University, Qingdao, 266237, China}

\author{S.~Tani}

\author{A.~Harasawa}

\author{Y.~Fukushima}

\author{H.~Tanaka}

\author{R.~Noguchi}

\author{T.~Iimori}
\affiliation{Institute for Solid State Physics (ISSP), The University of Tokyo, Kashiwa, Chiba 277-8581, Japan}

\author{K.~Yaji}
\affiliation{Research Center for Advanced Measurement and Characterization, National Institute for Materials Science (NIMS), Tsukuba, Ibaraki 305-0003, Japan}

\author{M.~Fujisawa}
\affiliation{Institute for Solid State Physics (ISSP), The University of Tokyo, Kashiwa, Chiba 277-8581, Japan}

\author{S.~Shin}
\affiliation{Office of University Professor, The University of Tokyo, Chiba 277-8581, Japan}

\author{F.~Komori}

\author{Y.~Kobayashi}
\affiliation{Institute for Solid State Physics (ISSP), The University of Tokyo, Kashiwa, Chiba 277-8581, Japan}

\author{Takeshi~Kondo*}
\email{kondo1215@issp.u-tokyo.ac.jp}
\affiliation{Institute for Solid State Physics (ISSP), The University of Tokyo, Kashiwa, Chiba 277-8581, Japan}
\affiliation{Trans-scale Quantum Science Institute, The University of Tokyo, Bunkyo-ku, Tokyo 113-0033, Japan}

\date{\today}

%
%%%%%%%%%%%%%%%%%%%%%%%%%%%%%%%%%%%%%%%%%%%%%%%%%%%%%%%%%%%%%%%%%%%%
\begin{abstract}                                                   %
    %%%%%%%%%%%%%%%%%%%%%%%%%%%%%%%%%%%%%%%%%%%%%%%%%%%%%%%%%
We describe a setup of time-, spin-, and angle-resolved photoemission spectroscopy (tr-SARPES) employing a 10.7-eV ($\lambda$=115.6~nm) pulse laser at 1-MHz repetition rate as a probe photon source. This equipment effectively combines technologies of a high-power Yb:fiber laser, ultraviolet-driven harmonic generation in Xe gas, and a SARPES apparatus equipped with very-low-energy-electron-diffraction (VLEED) spin detectors. A high repetition rate (1~MHz) of the probe laser allows experiments with the photoemission space-charge effects significantly reduced, despite a high flux of 10$^{13}$ photons/s on the sample. The relatively high photon energy (10.7~eV) also brings the capability of observing a wide momentum range that covers the entire Brillouin zone of many materials while ensuring high momentum resolution. The experimental setup overcomes a low efficiency of spin-resolved measurements, which gets even more severe for the pump-probed unoccupied states, and affords for investigating ultrafast electron and spin dynamics of modern quantum materials with energy and time resolutions of 25~meV and 360~fs, respectively.
\end{abstract}
\maketitle
\def\thefootnote{*}\footnotetext{These authors contributed equally to this work}%\def\thefootnote{\arabic{footnote}}
%
%
%%%%%%%%%%%%%%%%%%%%%%%%%%%%%%%%%%%%%%%%%%%%%%%%%%%%%%%%%%%%%%%%%%%%
% Introduction                                                     %
%%%%%%%%%%%%%%%%%%%%%%%%%%%%%%%%%%%%%%%%%%%%%%%%%%%%%%%%%%%%%%%%%%%%
%
\section{Introduction}\label{intro}
Understanding the temporal evolution not only of electrons in solids optically excited by ultrashort laser pulse but also its spin properties has grown desired as a technological trend toward optical control of electronic spin information, so-called opto-spintronics. The ultrafast spin dynamics have been so far studied mainly by macroscopic optical methods combined with a pump-probe approach using an ultrashort pulse laser at a femtosecond timescale, such as time-resolved Kerr-rotation spectroscopy~\cite{Beaurepaire1996-hl,Hsieh2011-ys} and circular absorption spectroscopy~\cite{Stamm2007-ur,Iyer2018-fh}.
On the other hand, time-, spin-, and angle-resolved photoemission emission spectroscopy (tr-SARPES) is a unique and powerful experimental technique, as it allows one to study these physics by directly observing the band structure of solids.
It is based on standard angle-resolved photoemission spectroscopy (ARPES)~\cite{Damascelli2003-lb,Sobota2021-dj} equipped with electron spin detectors~\cite{Kessler1969-jf,Okuda2017-ht}, directly probing the spin degree of freedom in addition to the electronic band structures~\cite{Okuda2013-wg}. In tr-SARPES, a pump-probe method is further combined, allowing one to trace the temporal evolution of both the photoexcited electronic populations and spin polarizations in the energy-momentum resolved band structure. It has, indeed, been applied for several studies of ultrafast dynamics in ferromagnets~\cite{Vaterlaus1991-pl,Scholl1997-ep,Aeschlimann1997-qe,Weinelt2007-ng,Winkelmann2008-sm,Cinchetti2009-fk,Schmidt2010-fd,Eich2017-rc,Buhlmann2020-nd,Andres2021-ez} as well as spin-orbit coupled materials~\cite{Cacho2015-vw,Jozwiak2016-tb,Sanchez-Barriga2016-aa,Fanciulli2020-yz,Clark2021-if,Mori2023-ut}. Nevertheless, it is still hard to insist that tr-SARPES has been established as a general-purpose experimental tool mainly because of its low efficiency in accumulating data, and thus the further development of this technique is strongly desired. 

In the past few decades, acute development was seen in tr-ARPES (without spin detection), and it remarkably boosted the research field for ultrafast dynamics in condensed matter physics; for example, this technique was employed to investigate the dynamics of photoexcited electronic states above the Fermi level ($E_{\rm{F}}$)~\cite{Echenique2004-kf,Bauer2005-ec,Bovensiepen2012-yq} and to control material properties~\cite{Gudde2007-jn,Wang2013-ob,Kuroda2016-ta,Mahmood2016-fs,Reimann2018-dr,Zhou2023-vi}. This situation arose due to continuous technological improvement not only of the electron analyzers but also of new laser sources. The tr-SARPES measurements have been conducted so far with a probe pulse laser either of $\sim$6~eV generated by non-linear optical crystals or of extreme ultraviolet (EUV) by high-harmonic generation (HHG) in rare gases ~\cite{Winterfeldt2008-to,Krausz_rmp2009}. In particular, HHG has recently been getting popular as a photon source of tr-ARPES. While deserting energy and momentum resolution, this brings the capability of observing a wide momentum space~\cite{Saule2019-wv,Chiang2012-nh,Emaury2015-ev,Corder2018-zq,Puppin2019-eq,Keunecke2020-ma,Suzuki2021-ru}, in contrast to a 6-eV pulse laser generally used, which cannot observe the entire Brillouin zone of solids.

\begin{figure*}[t!]
    \begin{center}
        \includegraphics[width=1.0\textwidth]{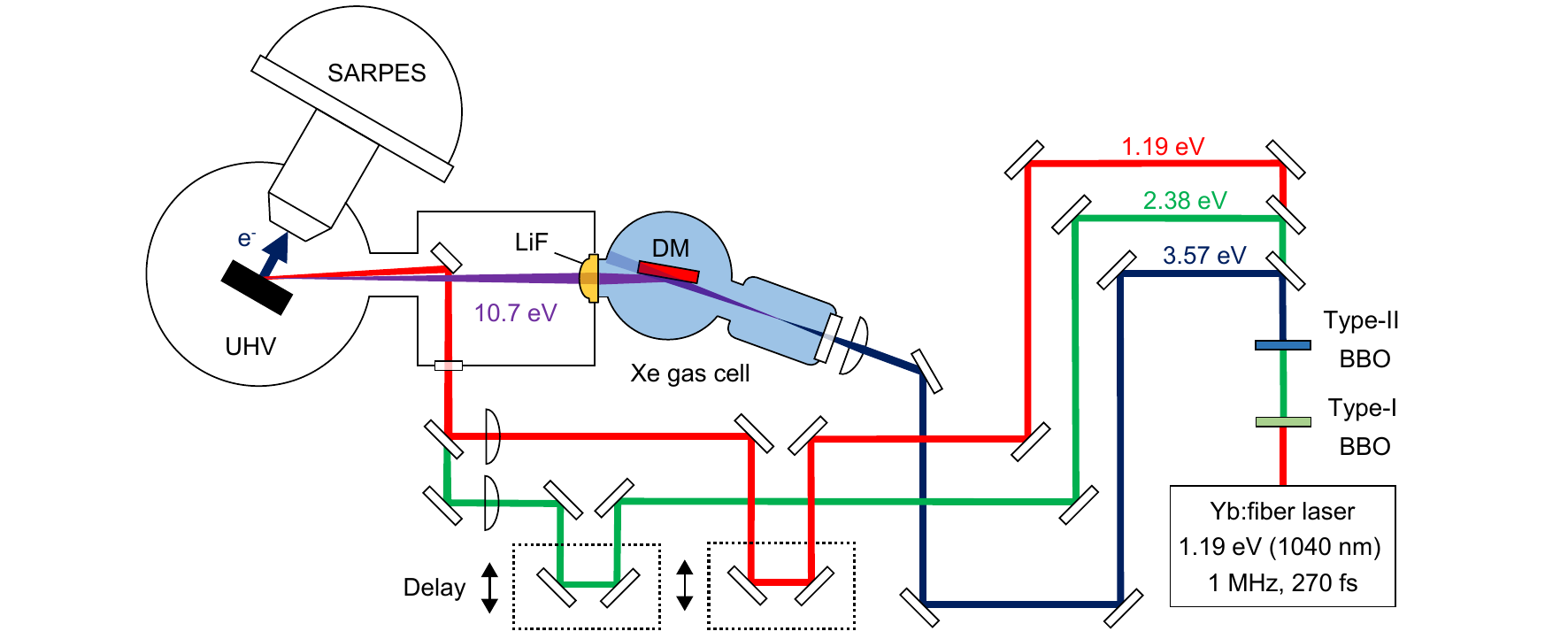}
        \caption[]{Schematic layout of the tr-SARPES system. A Yb:fiber laser (1.19~eV) is used as a high-power pulsed laser source at 1~MHz. Two $\beta$-BaB$_2$O$_4$ (BBO) crystals  are used for the second- and third-harmonic generations (SHG and THG, respectively). The THG pulses (3.57~eV) are focused onto a Xe gas cell to generate vacuum ultraviolet (VUV) pulses with a photon energy of 10.7-eV. The 10.7-eV pulses are reflected by a dichroic mirror (DM) in the Xe gas cell and focused by a lithium fluoride (LiF) lens-window onto a sample in an ultra-high vacuum (UHV) chamber equipped with a SARPES analyzer.
        }
        \label{fig:schem}
    \end{center}
\end{figure*}
\begin{figure*}[t!]
    \begin{center}
        \includegraphics[width=0.95\textwidth]{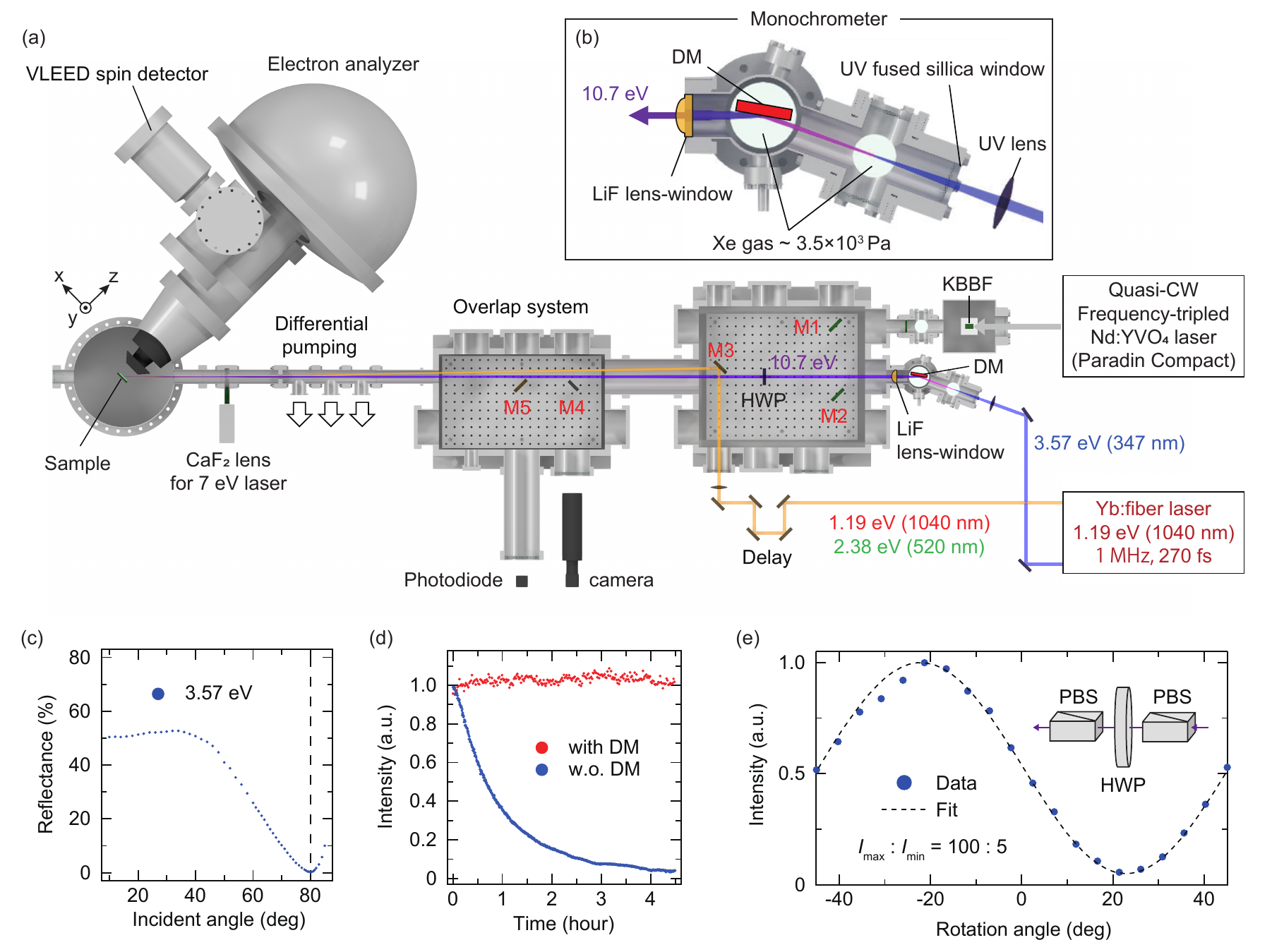}
        \caption[]{(a) Scaled layout of the 10.7-eV pulse laser beamline for tr-SARPES. M1, M2 : dielectric mirrors that are used for 7-eV laser, M3 : silver mirror for pump laser, M4 : silver mirror used for viewing a sample by a camera, and M5 : MgF$_2$ coated silver mirror used for adjusting the overlap between pump and probe pulses with a photodiode. The details of the setup are given in Sec.~\ref{beamline_setup}.
            (b) Magnified view of the Xe gas cell chamber including DM and the LiF lens-window. (c) Incident-angle dependence of DM reflectance of the UV-THG driver (3.57~eV). 
            (d) Intensity of the 10.7-eV pulse laser transmitted through the LiF lens-window with DM (red) and without DM (blue)~\cite{Zhao2019-mw}.
            (e)  The $p$-component intensity of 10.7-eV pulse laser that is polarization-converted by the MgF$_2$ half-wave plate (HWP) plotted as a function of the HWP rotation angle. For proper detection, the $p$-component is precisely selected by two Rochon prism beam splitters (PBS) placed in front of and behind HWP as shown in the inset.
        }
        \label{fig:beam}
    \end{center}
\end{figure*}

Despite the great successes of tr-ARPES, there are two strong limitations for further developing it to the spin-resolved version, tr-SARPES. The first limitation is that the spin detection efficiency is so low (typically only below $10^{-2}$ times the standard ARPES) and it gets even lower in the photoexcited occupied states that the measurement is very time-consuming~\cite{Okuda2017-ht}. This demands a large number of photons to acquire a sufficient count of photoelectrons. However, the number of photons per pulse cannot be increased since the dense population of photoelectrons excited at once by a pulse leads to space-charge effects which worsen energy and momentum resolutions~\cite{Zhou2005-as,Passlack2006-dn,Graf2010-bo,Oloff2016-qo,Ishida2016-lj}. This is the second limitation. Ironically, these two limitations are tied with each other, thus hardly solved at the same time. In particular, a strong pulse is, in principle, necessary for HHG at high photon energies. Hence, a reduced repetition rate is required, but it results in low efficiency, serious for tr-SARPES. While a 6-eV pulse laser has an advantage in gaining a high repetition rate, a problem is raised again, which is that the measurements are restricted to a narrow momentum region; here is another dilemma.

In this paper, we describe an approach to overcome the above limitations, realizing tr-SARPES measurements reconciling high efficiency in spin detection, a wide momentum space view, and good energy/momentum resolutions. The system combines the SARPES apparatus with a Yb:fiber laser system at the high-repetition rate 1~MHz with a high-power achieved by a chirped pulse amplification (CPA). A 10.7-eV probe pulses are generated as the third harmonic of the UV-THG driver (3.57~eV) in a Xe gas cell, providing a photon flux exceeding 10$^{13}$~photon/s on the sample at a repetition rate of 1 MHz. The VUV source combined with a SARPES end-station brings the capability of performing tr-SARPES experiments with a time resolution of 360~fs and an energy resolution of 25~meV corresponding to the laser bandwidth. 

The structure of the paper is as follows: Sec.~\ref{instrument_layout} overviews our apparatus. In Sec.~\ref{beamline_setup}, 
the tr-SARPES beamline will be introduced and characterized. We also present the overall performance of our tr-SARPES system by showing the experimental data for several materials in Sec.~\ref{performance}.

\section{Instrument layout}\label{instrument_layout}
Figure.~\ref{fig:schem} shows an overview of our tr-SARPES apparatus with an ultrashort 10.7-eV pulse laser, comprising three major sections: fundamental Yb:fiber laser source, Xe gas cell chamber, and SARPES apparatus. Combining these sections is crucial for the development of the state-of-the-art tr-SARPES apparatus. In the following, the details of each section will be described.

\subsection{1-MHz Yb:fiber laser source}\label{Yblaser_source}
An ideal light source for tr-SARPES needs to fulfill certain conditions regarding photon flux, repetition rate, and bandwidth. In contrast to standard photoemission experiments, tr-SARPES collects multi-dimensional data: the photoelectron intensity and the spin polarization vector as a function not only of energy and momentum but also of pump-probe delay time. Therefore, one has to accumulate the photoelectron signals from a large number of laser pulses to obtain data of sufficient statistics. However, the intensity of each probe pulse cannot be increased in tr-SARPES because the photoelectron clouds per pulse lead to strong space-charge effects~\cite{Zhou2005-as,Passlack2006-dn,Graf2010-bo,Oloff2016-qo,Ishida2016-lj}. The best way to mitigate it is to reduce the number of photons per pulse and instead increase the repetition rate of the pulse laser.

For this purpose, Yb:fiber laser could be a better light source than Ti:sapphire laser more commonly used. One of the advantages of Ti:sapphire laser is that a strong pulse with many photons can be generated. Therefore, this type of laser is widely used for the EUV generation (higher photon energy than 10~eV) at a kHz level. However, the EUV generation at higher repetition rates of an MHz level is still technical challenges~\cite{Saule2019-wv,Chiang2012-nh,Emaury2015-ev,Corder2018-zq,Puppin2019-eq,Keunecke2020-ma,Suzuki2021-ru}. In contrast, Yb:fiber system has advantages in that one can make the average power very high~\cite{Richardson2010-ju}. Recently, high pulse energy sufficient for the EUV generation even at a high repetition has been achieved by the Yb:fiber laser. In the present work, we design the instrument of tr-SARPES using a 10.7-eV probe laser generated by the home-built Yb:fiber CPA laser system operating at 1-MHz repetition rate~\cite{Zhao2015-th,Zhao2017-oa}.

\subsection{Design of Xe gas chamber}\label{source}
The vacuum ultraviolet (VUV) probe pulse laser of 10.7~eV ($\lambda$=115.6~nm) is generated via the frequency up-conversion in Xe gas~\cite{Hilbig1982-er,Zhao2017-oa,Peli2020-gm}. As a driver for it, we use
the ultraviolet laser (UV) at 3.57~eV produced by two BBO crystals as the third-harmonic generation (THG) of the Yb:fiber laser. Notably, the UV gives a good phase matching condition to achieve a high efficiency of 10$^{-4}$ even at a high-repetition-rate of the MHz-level~\cite{Zhao2019-mw}.

In our design for the beamline, the Xe gas-cell chamber plays key roles in two ways. One is for the 10.7-eV generation. The second is as a beam separation between the UV-THG driver and VUV pulse. Differently from the set-up for the 11-eV laser previously reported~\cite{Peli2020-gm,He2016-yo,Lee2020-wo}, we place an antireflection-coated mirror inside of our gas-cell chamber. It works as a dichroic mirror (DM) separating the UV and VUV lasers.

The gas-cell chamber is isolated from the vacuum chamber by a LiF window through which the generated 10.7-eV beam penetrates. LiF can be damaged by intense pulses in the UV range~\cite{Schneider1937-yb,Warneck1965-qp}, so the transmittance of the 10.7-eV beam would be degraded by the UV driver with the high-power shed on the LiF window. We avoid this issue by placing a DM in front of the LiF window. The idea of placing a DM in the gas cell chamber is the most important in our design of the 10.7-eV beamline for tr-SARPES.

\subsection{SARPES apparatus}\label{sarpes}
The above optical system is connected to the tr-SARPES end-station~\cite{Yaji2016-nn,Kuroda2018-yh}.
This SARPES system is equipped with the VLEED spin detectors~\cite{Okuda2017-ht}, which achieves the efficiency of 10$^{-2}$, 100 times larger than that of the Mott spin detectors~\cite{Okuda2017-ht}. 
The data quality, therefore, can be much better and the acquisition time gets much shorter. The combination of the high-efficient SARPES end station equipped with the VLEED spin detectors and the photon source of the 10.7-eV pulse laser at the high repetition rate (1~MHz) is a crucial element for realizing our state-of-the-art tr-SARPES apparatus.

\section{Beamline setup and specification}\label{beamline_setup}
%\section{III. Setup and specification}
\subsection{Overview}
A scaled layout of the pump-probe beamline with the 10.7-eV pulse laser constructed for tr-SARPES is sketched in Fig.~\ref{fig:beam}. As a fundamental light source, we use a home-built Yb:fiber CPA laser set at 1~MHz~\cite{Zhao2017-oa}.
It produces 270~fs pulses at 1.19~eV and high-power up to 100~$\mu$J/pulse.
The whole fiber laser system is placed in a laser booth where temperature and humidity are stabilized.

The UV pulse laser at 3.57~eV produced by THG via BBOs is focused onto our Xe gas cell chamber by a UV lens with $f$=150 mm (Throlabs, LA4874-UV), generating the 10.7~eV pulses. We use a beam stabilizer (TEM Messtechnik GmbH, Aligna) to compensate for drifts of the UV path before reaching the laser table outside of the laser booth.

In the Xe gas-cell chamber (inset of Fig.~\ref{fig:beam}), the generated 10.7-eV pulses are reflected by DM (NTT-AT corp.) at the gracing angle of 10~$^{\circ}$ to satisfy the antireflection condition for the $p$-polarized UV-THG driver as shown in Fig.~\ref{fig:beam}(c). DM is made of the multilayer coating of SiO$_2$/TiO$_2$ on a fused silica substrate and terminated by a TiO$_2$ layer on the surface to obtain a high reflection of 10.7~eV. The reflection of 30~$\%$ is expected by calculations with the optical constant of TiO$_2$.

%This mirror is made of a multilayer coating of SiO$_2$/TiO$_2$ on a fused silica substrate and terminated by a TiO$_2$ layer on the surface to give a good reflection (30~$\%$) of 10.7~eV.

The pressure of the inner Xe gas, sealed by the LiF lens-window (Korth Kristalle GmbH) and the fused silica window (Thorlabs, WG42012-UV), is kept at 3.5$\times$10$^3$~{Pa} to obtain a high efficiency of the 10.7-eV generation.
The LiF lens-window is integrated as a piece with a convex lens ($f$=150~mm), mildly focusing the 10.7-eV laser onto the sample in the SARPES chamber.  Importantly, this LiF lens-window also acts as an "anti-achromatic" lens for the UV-THG driver and 10.7-eV pulse lasers: Because the refractive index of LiF for their wavelength is different, only the 10.7-eV laser is focused onto the samples. Thanks to this property, the density of UV-THG driver slightly reflected by DM ($\sim$0.1~$\%$ reflection) becomes very small at the sample position to be less than 1~pJ/cm$^2$, which is negligible for pump-probe experiments.

Degradation of the LiF transmission by the irradiation of the UV pulses \cite{Schneider1937-yb,Warneck1965-qp} is also avoided  by using DM. Figure~\ref{fig:beam}(c) presents the intensities of the transmitted 10.7-eV laser from the LiF lens-window under two different settings monitored by the phototube (Hamamatsu Corp., R1187) [Fig.~\ref{fig:beam}(d)]: one is with DM as mentioned above, and the other is without DM but instead uses a LiF prism to work as a separator of 10.7-eV pulses from the UV-THG driver laser~\cite{Zhao2019-mw}. Without DM, the LiF lens-window is irradiated by the bright UV-THG pulses, showing rapid degradation in the transmittance of the 10.7-eV laser [blue marks in Fig.~\ref{fig:beam}(d)]. In contrast, the transmittance becomes stable (red marks) when using DM, indicating the LiF lens-window is protected from irradiation damage.

One of the advantages of using the 10.7-eV pulse laser as the probe is that one can easily control the light polarization with a birefringence crystal, MgF$_2$. The evaluation of the half-waveplate (HWP) of MgF$_2$ (Kogakugiken Corp.) is presented in Fig.~\ref{fig:beam}(e), where the intensity of the transmitted 10.7~eV pulse laser detected by the phototube is plotted as a function of HWP angle. For this test measurement, only the $p$-polarization component is detected (see the inset).
By changing the HWP angle from 0$^\circ$ to 45$^\circ$, the transmitted intensity draws a sinusoidal curve, confirming that light polarization changes from $p$ to $s$. The polarization extinction ratio is estimated to be 100:5, which is sufficient for studying the light polarization dependence of photoelectron spin-polarization~\cite{miyamoto_prb16,Yaji2018-ds}.

The beamline is connected to the SARPES chamber~\cite{Yaji2016-nn,Kuroda2018-yh} with differential pumping systems keeping the UHV condition in the order of 10$^{-9}$~Pa. The hemispherical electron analyzer is ScientaOmicron DA30L equipped with double VLEED spin detectors, which permit a vector analysis of photoelectron spin-polarizations in three dimensions~\cite{Okuda2015-ug}.  A molecular beam epitaxy (MBE) chamber is connected to this apparatus, and grown thin films can be transferred $in\; situ$ to the measurement chamber.

In front of the SARPES chamber, a metallic mirror (M5) is placed. This is mounted on a linear translation stage and can be inserted in the beam path to reflect the pump pulse and the residual UV-THG driver passing through the Xe chamber. 
This setup allows one to adjust for the pump and probe pulses to have a temporal overlap while checked by the photodiode (EOT Corp., ET-2030). For this aim, the light polarization of the UV-THG driver needs to be changed to $s$-polarization as being reflected at DM and detected with the photodiode. In order to achieve a spatial overlap of the pump and probe beams, we use a pinhole of polycrystalline gold (Au) films deposited on an oxygen-free copper plate~\cite{Ishida2014-jv}. The 10.7~eV probe beam is aligned to the pinhole position by manually controlling a mirror mount in front of the gas cell chamber as monitoring with a long-focus camera via M4. Once the 10.7-eV beam is fixed at the pinhole, the pump beam position is optimized, while observed by a camera, by controlling a motorized mirror mount of M3.

In addition to the set-up for the pump-probe tr-SARPES, the beamline also connects a chamber installing KBe$_2$BO$_3$F$_2$ (KBBF), which is isolated by a CaF$_2$ window.  This enables measurements of the high-resolution SARPES~\cite{Yaji2016-nn,Kuroda2018-yh} with the 7-eV laser, a second-harmonic of the frequency-tripled Nd:YVO$_4$ laser~\cite{Shimojima2015-hd}. The 7-eV laser is carried to the SARPES chamber by two dielectric mirrors (M1 and M2) and focused onto the samples by the CaF$_2$ lens (Kogaku Corp.) mounted on the UHV gate valve. The linear translation stage carrying M2 is motorized, which enables switching the light source between 10.7~eV and 7~eV for the measurements.

\begin{figure}[t!]
    \begin{center}
        \includegraphics[width=1.0\columnwidth]{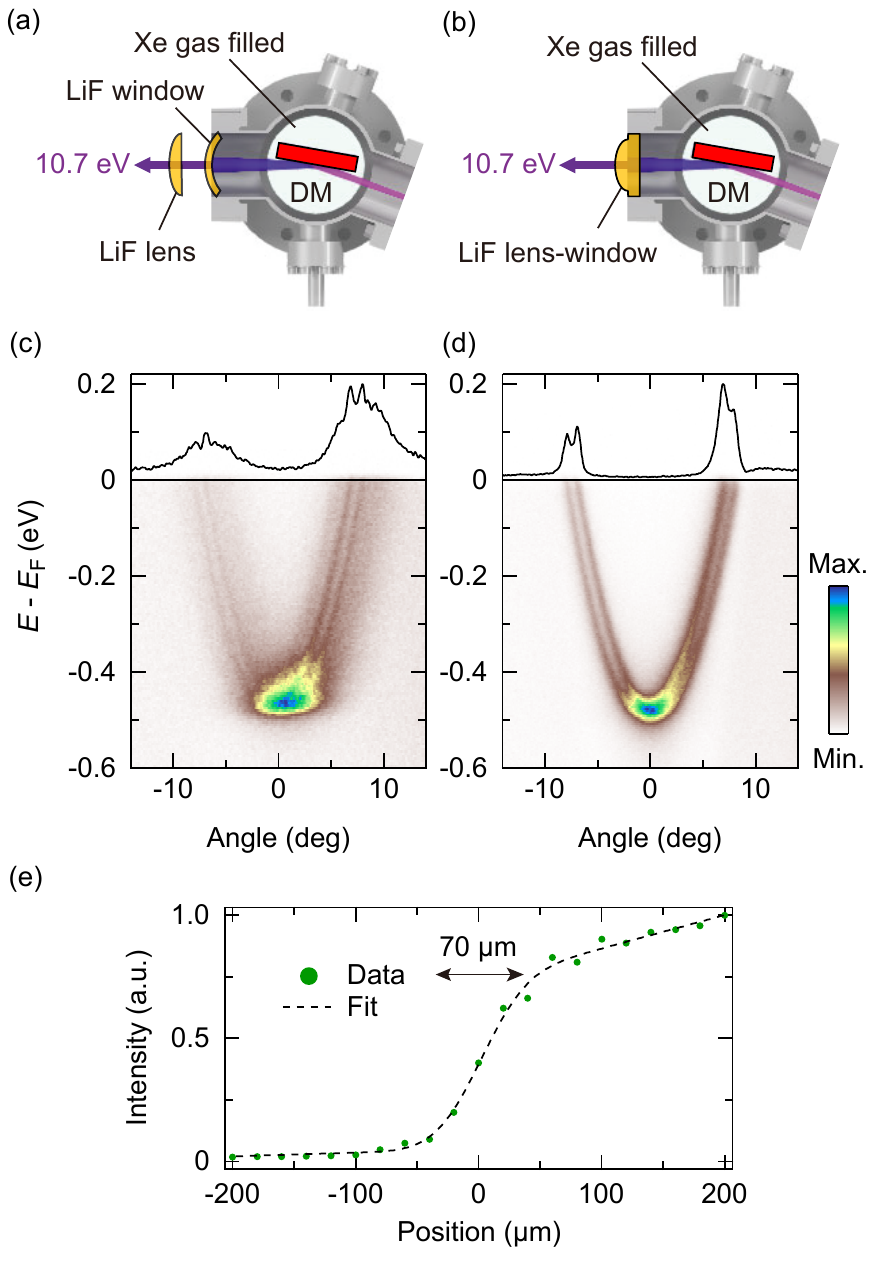}
        \caption[]{Two different setups of LiF lens and window for the Xe gas cell. (a) LiF window and lens are separated. (b) These two are integrated as a piece (LiF lents-window). 
            (c,d) ARPES results of Shockley surface state at (111) surface of gold single crystal and the momentum distribution curves at Fermi energy, measured at $T$=28~K for the setups of (a,b), respectively.
            (e) Angle-integrated photoemission signals obtained in the lens-window setup from the knife edge of polycrystalline gold films deposited on an oxygen-free copper plate. The spot size of the 10.7-eV pulse laser at the sample position is estimated to be 70~$\mu$m from the full width at half maximum.}
        \label{fig:LiF}
    \end{center}
\end{figure}
\begin{figure}[t!]
    \begin{center}
        \includegraphics[width=1.0\columnwidth]{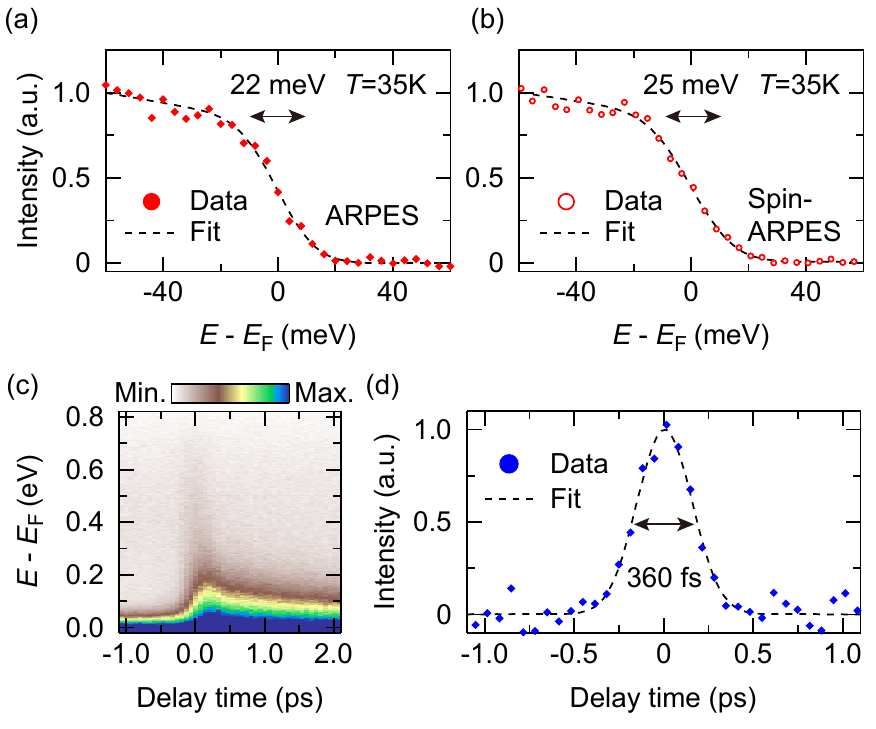}
        \caption[]{Characterization of the energy and time resolutions.
            (a) ARPES spectrum for an evaporated polycrystalline bismuth films measured at $T$=35~K. The fitting curve (dashed line) estimates the energy resolution to be 22 meV.  
            (b) Spin-ARPES spectrum for the same polycrystalline bismuth films as used in (a), detected by the VLEED spin detector~\cite{Yaji2016-nn}. The fitting curve (dashed line) estimates the energy resolution to be 25 meV.  
            (c) Pump-probed signal of highly oriented pyrolytic graphite (HOPG) acquired at $T$=83~K. Here, the intensities of angle-integrated photoelectrons around $\Gamma$ are presented. The 1.19-eV pulse is used for the pump. 
            (d) Time evolution of pump-probed signals at $E$$-$$E_{\rm{F}}$=0.6$\pm$0.1~eV extracted from (d). 
            The experimental time resolution is estimated to be 360 fs from the full width at half maximum of the Gauss function fitted to the data (dashed line). This resolution corresponds to the time duration of a temporal cross-correlation between the pump and probe.}
        \label{fig:resolution}
    \end{center}
\end{figure}

\begin{figure}[t]
    \begin{center}
        \includegraphics[width=1.0\columnwidth]{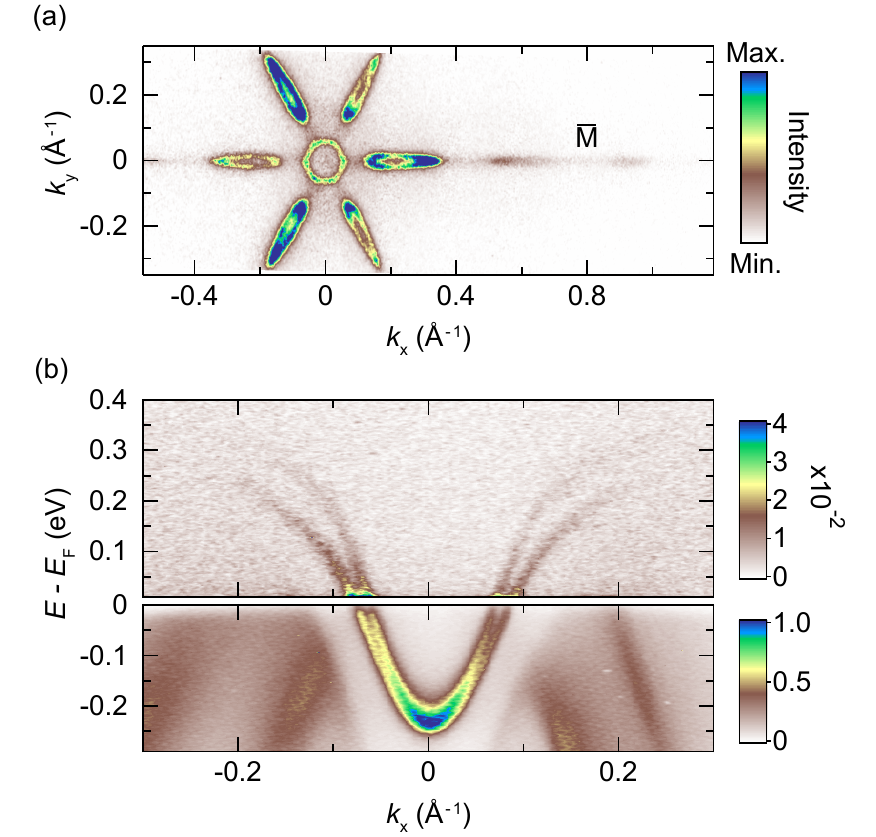}
        \caption[]{(a) The Fermi surfaces of bismuth thin films grown on a Si(111) substrate obtained at $T$=50~K. The data were taken with the sample rotation and using an electron deflector equipped in the photoelectron analyzer~\cite{Yaji2016-nn}.
        (b) Pump-probed ARPES image of the surface state along $\bar{\Gamma}$-$\bar{\mathrm{M}}$ on the (111) face of grey arsenic at 300~fs after excitation by pump pulse (1.19~eV). The data were measured at $T$=18~K. The color scale of the upper image is changed from that of the lower image, to increase the visibility of the unoccupied states above $E_{\rm{F}}$ more clearly.}
        \label{fig:Bi}
    \end{center}
\end{figure}

\subsection{Specifications}

We present the performance of the beamline of the 10.7-eV pulse laser; photon flux, spot size, and energy/time resolutions.
To obtain a good performance of these, the LiF lens-window used at the Xe gas cell chamber needs to be carefully designed.
The transmittance of LiF crystal for the 10.7~eV is only 40~$\%$ mainly due to Fresnel reflection. Thus, one should reduce the number of LiF optics in the overall beamline to secure the photon flux of the transmitted beam. The photon flux gets higher with the all-in-one type LiF lens-window [Fig.~\ref{fig:LiF}(b)] than in the case when the LiF window and lens are separated [Fig.~\ref{fig:LiF}(a)]. In our beamline [Fig.~\ref{fig:beam}], by using this LiF lens-window [Fig.~\ref{fig:LiF}(b)], we achieve to obtain a high photon flux $\sim$2.6$\times$10$^{13}$ photon/s (corresponding to $\sim$50~$\mu$W) of the 10.7-eV pulses
at the sample position in the SARPES chamber. The photon flux was determined by the phototube (Hamamatsu corp., R1187). To achieve this high photon flux, we use 10~W of the fundamental 1040~nm and the accordingly generated 1~W of UV-THG driver.

\begin{figure*}[t]
    \begin{center}
        \includegraphics[width=0.92\textwidth]{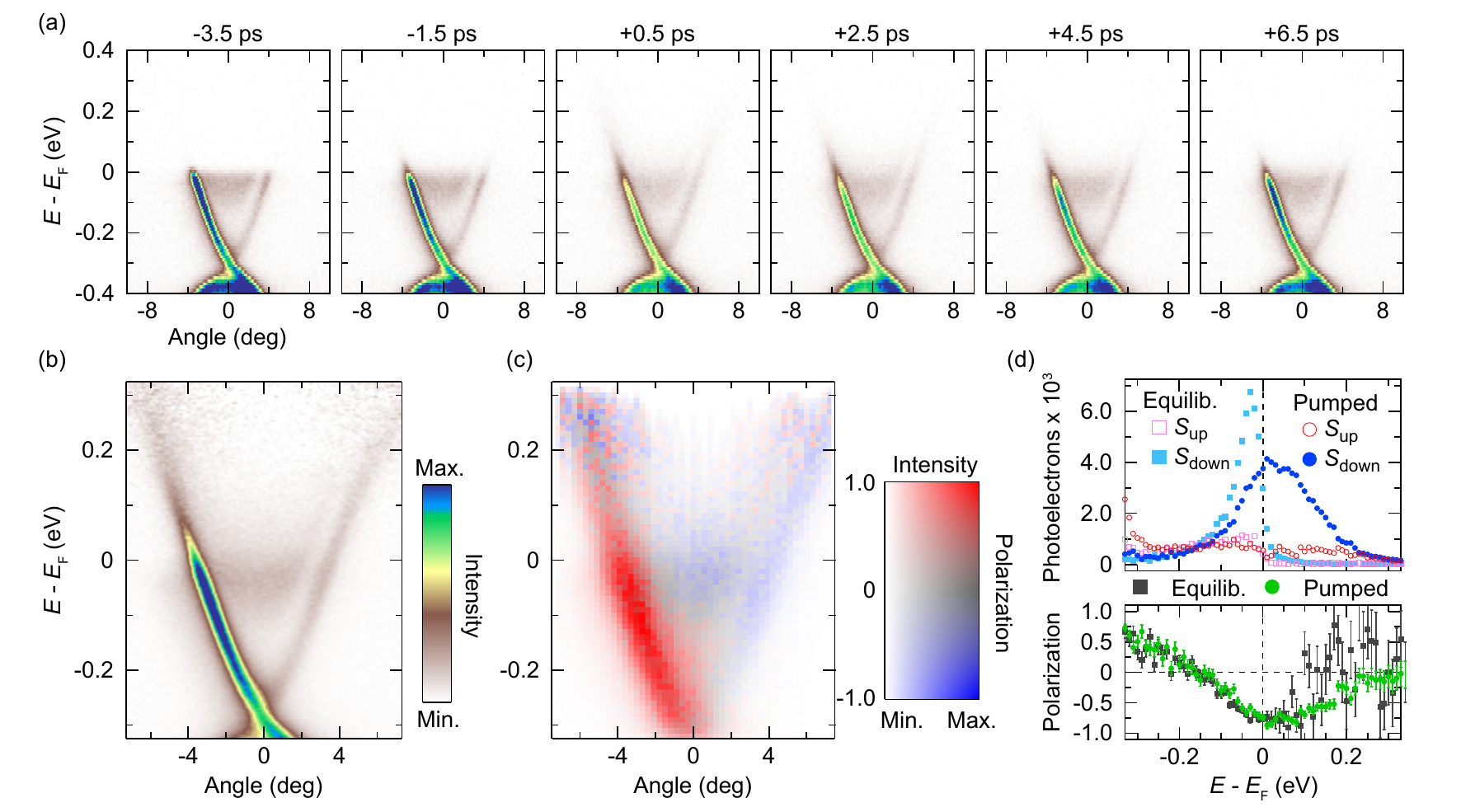}
        \caption[]{(a) The tr-ARPES images of Bi$_2$Se$_3$ at various delay times between the pump (1.19~eV) and probe (10.7~eV) pulses. The data were measured at $T$=23~K, and each map was obtained within 5~min.
            (b) Representative ARPES map at 0.5~ps after the pump. The image is divided by the Fermi-Dirac function at an electronic temperature ($k_{\rm{B}}T$=60~meV) to enhance the visibility above $E_{\rm{F}}$.          
            (c) The spin polarization intensity map corresponding to (b), expressed with two-dimensional color code~\cite{Yaji2018-ds}. This SARPES image was obtained within 8~hours.
            (d) SARPES spectra (upper panel) and the corresponding spin-polarization spectra (lower panel) taken at $+$4~$^{\circ}$ without (squares) and with (circles) pump excitation. The accumulation times for each data set without/with the pump are  30~min.
        }
        \label{fig:Bi2Se3}
    \end{center}
\end{figure*}

LiF has ionic crystallinity and is, therefore, so weak against physical stress that the LiF window could be easily bent by the Xe gas pressure. In fact, we observe such a distortion occurring for the LiF window thinner than 2~mm.
This distortion in the window induces light scattering and hence worsens the beam focus. It is noticeable in ARPES signals demonstrated in Fig.~\ref{fig:LiF}: poor quality of a probe laser affected by a LiF thin window [Fig.~\ref{fig:LiF}(a)] causes the broadening of the ARPES spectra for the Shockley surface state of Au(111)~\cite{LaShell1996-ni} [Fig.~\ref{fig:LiF}(c)]. The beam quality is improved by using the LiF lens-window of 2~mm thickness, and consequently, ARPES spectra become high quality [Fig.~\ref{fig:LiF}(d)].  In Fig.~\ref{fig:LiF}(e), we examine the beam profile for the setup using the LiF lens-window [Fig.~\ref{fig:LiF}(b)] by plotting photoemission intensities for a knife edge of a polycrystalline gold films deposited on an oxygen-free copper plate. The beam spot size is estimated to be 70~$\mu$m in the full width at half maximum. 

In order to evaluate the energy resolution of ARPES and SARPES with the 10.7-eV pulse laser, we measure a polycrystalline bismuth films deposited on an oxygen-free copper plate [Fig.~\ref{fig:resolution}(a) and (b)]. The angle-integrated energy distribution curves (EDCs) were fit with a Fermi-Dirac function convoluted with a Gaussian function (black lines). The instrumental energy resolution ($\Delta{E}_{\rm{inst}}$) is determined to be 22~meV from the width of the obtained Gaussian function. This resolution is expressed as $\Delta{E}_{\rm{inst}}$=$\sqrt{\Delta{E}_{\rm{ana}}^2+\Delta{E}_{\rm{probe}}^2}$ with the energy resolution of the electron analyzer ($\Delta{E}_{\rm{ana}}$) and a bandwidth of the 10.7~eV laser pulse ($\Delta{E}_{\rm{probe}}$). $\Delta{E}_{\rm{ana}}$ is expected to be rather small ($\sim$3~meV) according to $\Delta{E}_{\rm{ana}}$= $E_{p}w$/2$R$: $R$ is the radius of the analyzer (200~mm), $E_{p}$ is the pass energy (5~eV), and $w$ is the entrance slit (0.2~mm). 
Therefore,  $\Delta{E}_{\rm{probe}}$ and the corresponding band width $\Delta\lambda$ are determined as 22 meV and 0.24 nm respectively.

To determine the energy resolution for SARPES, we measure the same bismuth films used for APRES [Fig.~\ref{fig:resolution}(b)].
Compared to ARPES [Fig.~\ref{fig:resolution}(a)], the photoelectron count rate is much lower in SARPES, since this technique detects photoelectrons reflected (or spin-resolved) by a VLEED spin target. Note that the VLEED spin detector is a highly efficient spin detector, but still, the reflection probability of the target is only $\sim$10~$\%$~\cite{Okuda2017-ht}. To compensate for the reduced count rates, we increase the analyzer slit up to 0.8~mm and select an aperture size of 
3$\times$0.5~mm~\cite{Yaji2016-nn} for spin detection. These, however, do not worsen the energy resolution much. 
$\Delta{E}_{\rm{inst}}$ in SARPES is estimated to be 25~meV [Fig.~\ref{fig:resolution}(b)], which is comparable to that of ARPES. Even in SARPES, $\Delta{E}_{\rm{probe}}$ is the primal factor deciding the energy resolution. 

The Fourier transform limited pulse duration ($\Delta t_{\rm{probe}}$) of the bandwidth ($\Delta{E}_{\rm{probe}}$=22~meV) is estimated as 83~fs according to the following equation:$\Delta E_{\rm{probe}} \Delta t_{\rm{probe}}$=$4 \hbar \ln{2}$ ($\hbar$ is reduced Planck constant).
However, LiF crystals used for the lens-window could cause a chirp for the 10.7-eV pulse, decreasing the experimental time resolution. To determine the resolution including the chirp effect, we observe in Fig.~\ref{fig:resolution}(c) the pump-probed signal of HOPG (a cleaved surface) which provides a fast response limited by the experimental time resolution~\cite{Ishida2014-jv}. 
Excited electron signals are seen in a very short time around $t$=0~ps when the overlap of the pump and probe pulses is large. 
The temporal profile at $\sim$0.6~eV above $E_{\rm{F}}$ shows a Gaussian shape [Fig.~\ref{fig:resolution}(d)], verifying that the fast response is resolution-limited. By fitting the profile with a Gaussian function, the time resolution ($\Delta{t}$) is estimated to be 360~fs. Since the duration of the pump pulse (1.19 eV), determined by an autocorrelation method, is 270~fs, that of the 10.7-eV pulse duration is estimated as 230~fs.

%
%
%%%%%%%%%%%%%%%%%%%%%%%%%%%%%%%%%%%%%%%%%%%%%%%%%%%%%%%%%%%%%%%%%%%%
% Results                                                          %
%%%%%%%%%%%%%%%%%%%%%%%%%%%%%%%%%%%%%%%%%%%%%%%%%%%%%%%%%%%%%%%%%%%%
%
%
%
%
\begin{figure*}[t]
    \begin{center}
        \includegraphics[width=0.95\textwidth]{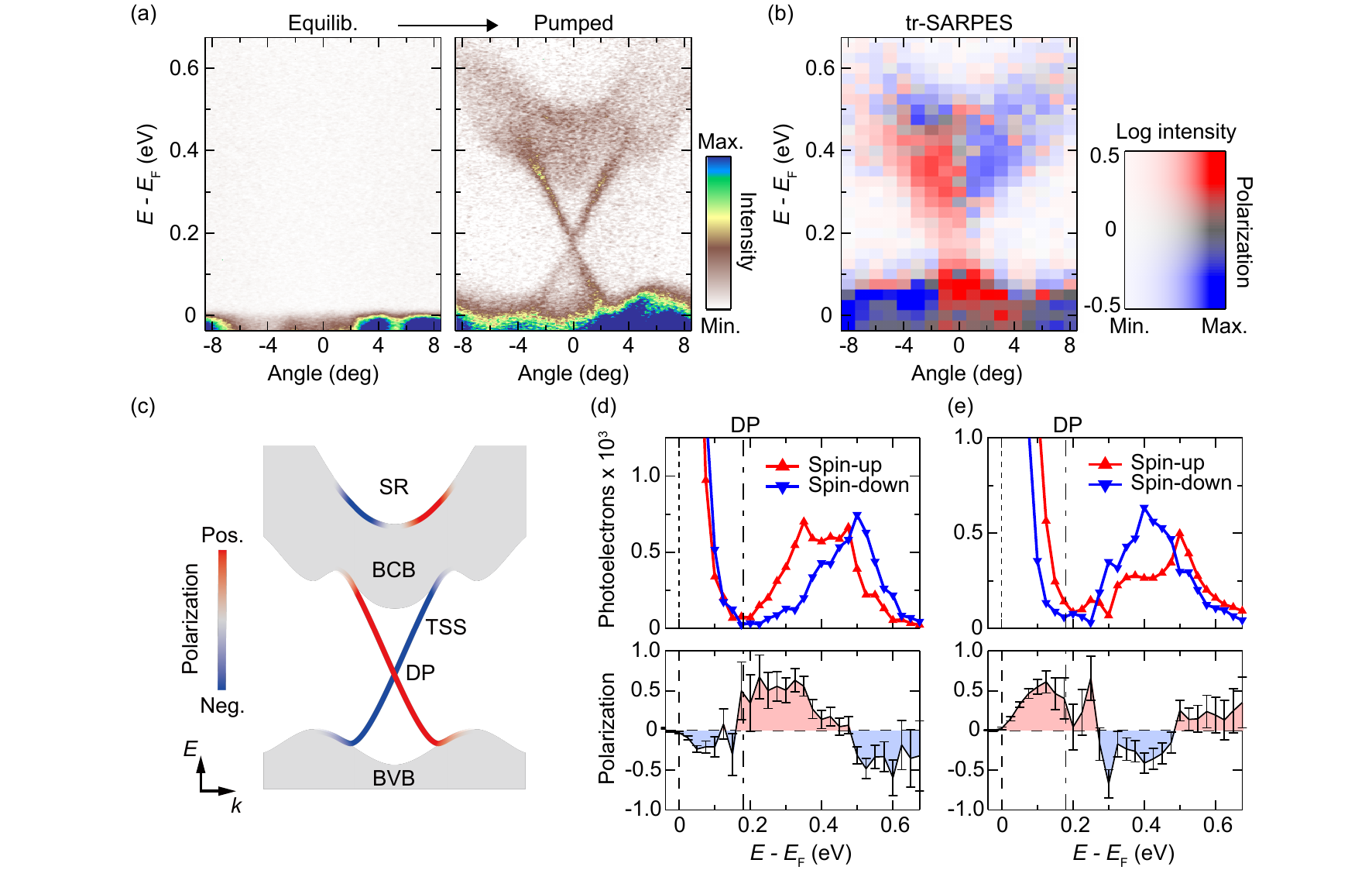}
        \caption[]{(a) Pump-probe ARPES image of Sb$_2$Te$_3$ without pump (the left panel) and at 0.5~ps after the pump (the right panel; $\hbar\omega_{\rm{pump}}$=1.19~eV). (b) The tr-SARPES image of Sb$_2$Te$_3$ at 500~fs after the arrival of the pump pulse. 
 The data were measured at $T$=25~K. (c) Schematic of our tr-SARPES results illustrating bulk valence and conduction bands (BVB and BCB), topological surface state (TSS), Dirac point (DP), and unoccupied surface resonance (SR)~\cite{Jozwiak2016-tb}. TSS has a spin-helical Dirac cone, and SR is also spin-polarized.  
BCB and BVB show continuum states projected on the surface (shaded area). (d,e) Spin-resolved EDCs (upper panel) and the corresponding spin-polarization (lower panel) at $-$3~$^{\circ}$ and $+$3~$^{\circ}$, respectively.}
        \label{fig:Sb2Te3}
    \end{center}
\end{figure*}
\section{Performance}\label{performance}

\subsection{Bismuth thin films and grey arsenic: wide $k$ maps}
An advantage of using a 10.7-eV laser is that one can access electronic structures over a wide momentum-range while keeping a high momentum resolution. Here, we demonstrate it by the ARPES measurements on bismuth that is known to host a metallic surface state on the (111) surface dispersing over the entire surface Brillouin zone~\cite{Hofmann2006-tw}.
The bismuth thin films~\cite{Nagao2004-bz,Hirahara2006-fh} is grown $\sim$100~bilayers $in\; situ$ on a $n$-type Si(111) substrate in the MBE chamber, and it is transferred to the SARPES chamber.
Figure~\ref{fig:Bi}(a) represents the Fermi surface map obtained by the 10.7-eV laser.
% \cite{He2016-yo,Peli2020-gm}
Thanks to high photon energy, one can observe the in-plane dispersion up to the zone boundary at $\bar{\mathrm{M}}$ ($k_{x}$$\simeq$0.8~\AA$^{-1}$), which cannot be accessed by a conventional laser such as 6-eV~\cite{Ishida2014-jv} and 7-eV~\cite{Yaji2016-nn} laser, and even far beyond it over a wider momentum space.

The capability of observing a wide momentum space with a 10.7-eV laser \cite{He2016-yo,Lee2020-wo} is also very useful in pump-probe experiments. In Fig.~\ref{fig:Bi}(b), we demonstrate it by measuring a cleaved (111) surface of a grey arsenic single crystal \cite{Zhang2017-bo}.  A nearly free-electron parabolic band for the Shockley surface state is observed around $\bar\Gamma$ on the occupied side below $E_{\rm{F}}$ [Fig.~\ref{fig:Bi}(b)]. When excited by the intense femtosecond pump pulses (1.19~eV, 0.1~$\mu$J/pulse), electrons are redistributed into the states above $E_{\rm{F}}$, imaging the unoccupied part of the surface band in a wide momentum view. It is clearly exhibited that the band dispersion deviates from the simple parabolic shape by going away from $\bar\Gamma$ and eventually disappears by merging into the bulk continuum states.

\subsection{Bi$_2$Se$_3$: tr-SARPES}
Next, we perform tr-SARPES for a prototype topological insulator Bi$_2$Se$_3$ with a spin-polarized metallic surface state forming a Dirac cone on the (111) surface [Fig.~\ref{fig:Bi2Se3}]~\cite{Zhang2009-jn,Xia2009-gb}.
For this experiment, a clean (111) surface of a bulk Bi$_2$Se$_3$ crystal is prepared by cleavage in the UHV chamber. Figure~\ref{fig:Bi2Se3}(a) plots typical tr-ARPES images taken at several delay times between the pump (1.19~eV, 0.12~$\mu$J/pulse) and probe pulses. One can trace ultrafast carrier dynamics of photoexcited surface bands evolving within sub-picoseconds. The Dirac dispersion above $E_{\rm{F}}$ is populated just after the pump excitation, and the populated states gradually decay after a few picoseconds.

Now, we turn the measurements to the spin-resolved mode to show the capability of our tr-SARPES setup. 
Figures~\ref{fig:Bi2Se3}(b) and (c) present the pump-probed ARPES image and the corresponding spin polarization at 0.5~ps after the arrival of the pump pulse. To clearly show the band dispersions above $E_{\rm{F}}$, these spectral images are divided by the Fermi-Dirac distribution at electronic temperature ($T_{\rm{el}}$=700~K, $k_{\rm{B}}T$=60~meV) transiently increased by the pump excitation. The spin-polarization map [Fig.~\ref{fig:Bi2Se3}(c)] contains 66$\times$52 energy-angle points measured at the energy and angular steps of 10 meV and 0.3~$^{\circ}$, respectively. 
The energy and angular resolutions were each set to be 25~meV and 2.8~$^{\circ}$, and the acquisition times of the overall image were 8~hours. 

In Fig.~\ref{fig:Bi2Se3}(d), we plot the spin-resolved EDC (the top panel) and the corresponding spin-polarization (the bottom panel) measured at the emission angle $\theta$=4.2~$^{\circ}$. In these panels, the data taken with and without optical excitation by the pump pulse are overlayed to be compared. In the equilibrium state without pumping (opened marks), the spectral intensity above $E_{\rm{F}}$ is negligibly small because of the Fermi-energy cutoff. Although the spin-polarization is detected slightly above $E_{\rm{F}}$, it is due to the thermally excited electrons and limited up to $E$$-$$E_{\rm{F}}\sim$0.1~eV, corresponding to the measurement temperature ($T$=20~K). On the other hand, the spectral weight above $E_{\rm{F}}$ becomes dominant in the pump-probed data owing to transient populations excited up to higher binding energies. This leads to visualizing the spin-polarized band on the unoccupied side [Fig.~\ref{fig:Bi2Se3}(c)], which cannot be accessed by a conventional SARPES.

\subsection{Sb$_2$Te$_3$: spin map of the transient population}
We perform tr-SARPES for another topological insulator Sb$_2$Te$_3$.
This compound has a Dirac cone expected for a spin-polarized topological surface state. In contrast to Bi$_2$Se$_3$, however, the crystal of Sb$_2$Te$_3$ has naturally $p$-type characters~\cite{Ando2013-lz,Dai2016-so,Jiang2012-fn}. Hence, most part of the Dirac dispersion is located above $E_{\rm{F}}$. It has been indeed visualized by tr-ARPES~\cite{Reimann2014-zg,Zhu2015-lu}. However, the spin-polarization expected for the topological surface state has not been experimentally revealed. Our tr-SARPES is capable of performing it for the first time. 

Figures~\ref{fig:Sb2Te3}(a) and (b) show the band maps without and with pumping (1.19~eV, 0.2~$\mu$J/pulse). The pump-probed data is recorded at 0.5~ps after the arrival of the pump pulse when the intensity of the excited electrons reaches maximum~\cite{Reimann2014-zg,Zhu2015-lu}. It is demonstrated that the Dirac cone hidden on the unoccupied side at the equilibrium state entirely emerges by the optical excitation. From the data, one can identify the Dirac point to locate at $\sim$180~meV above $E_{\rm{F}}$ and the upper Dirac cone to be absorbed into the continuum of the bulk states around 0.4~eV.

By switching to the spin-detection mode, here, we present the first visualization of the spin-polarization for the surface Dirac cone of Sb$_2$Te$_3$, unraveling its helical spin texture~\cite{Zhang2009-jn}.
Figures~\ref{fig:Sb2Te3}(d) and (e) plot the spin-resolved EDCs (the upper panels) at fixed emission angles of $\pm$3~$^{\circ}$ and the corresponding spin-polarization (the lower panels). 
The data clearly shows a sign inversion of the spin polarization about the normal emission angle ($\theta$=0~$^{\circ}$) at $E$$-$$E_{\rm{F}}$=0.3~eV and 0.1~eV each corresponding to the lower and upper Dirac cones. 
Furthermore, the spin direction is opposite between the lower and upper Dirac cones. 
These results represent helical spin textures characteristic of the topological surface states, which flip direction 
across the Dirac point [see Fig.~\ref{fig:Sb2Te3}(c)]. Our data of spin polarization also show another sign inversion around $E$$-$$E_{\rm{F}}$=0.6~eV. This is most likely originating from a spin-polarized surface resonance state, as schematically depicted in Fig.~\ref{fig:Sb2Te3}(c), which is compatible with that recently observed in Bi$_2$Se$_3$~\cite{Jozwiak2016-tb}. Our new tr-SARPES equipment, therefore, entirely unveils the unoccupied spin-polarized states in Sb$_2$Te$_3$ for the first time.

%
%%%%%%%%%%%%%%%%%%%%%%%%%%%%%%%%%%%%%%%%%%%%%%%%%%%%%%%%%%%%%%%%%%%%
\section{Conclusion}
In conclusion, we introduced a setup for tr-SARPES apparatus with the 10.7-eV pulse laser at 1-MHz repetition rate 
capable of band mapping over a wide momentum space at a high momentum resolution. The light source is generated by the frequency up-conversion to 10.7~eV by the UV-THG driver and Xe-plasma, based on the Yb:fiber CPA laser system operating at 1~MHz. Combining this laser source with high-efficient VLEED spin detectors, tr-SARPES is feasible at the energy resolution of 25 meV and the time resolution of 360~fs. This setup can map the band structure of photoexcited states together with the spin polarization information and follow its temporal evolution on a femtosecond time scale. The newly developed tr-SARPES apparatus will be a powerful experimental tool for future studies of spin-polarized electronic states in modern quantum materials.

% Acknowledgements                                                 %
%%%%%%%%%%%%%%%%%%%%%%%%%%%%%%%%%%%%%%%%%%%%%%%%%%%%%%%%%%%%%%%%%%%%
%
\section{Acknowledgments}
We acknowledge Federico Cilento, Yukiaki Ishida, Suguru Ito for fruitful comments and discussions, and D.~Matsumaru, S.~Sakuragi T.~Suzuki, T.~Kurihara for supports of experiments. This work was supported by the JSPS KAKENHI (Grants Numbers. JP21H04439), by MEXT Q-LEAP (Grant No. JPMXS0118068681), and by MEXT as “Program for Promoting Researches on the Supercomputer Fugaku” (Basic Science for Emergence and Functionality in Quantum Matter Innovative Strongly-Correlated Electron Science by Integration of “Fugaku” and Frontier Experiments, JPMXP1020200104) (Project ID: hp200132/hp210163/hp220166).
%
%%%%%%%%%%%%%%%%%%%%%%%%%%%%%%%%%%%%%%%%%%%%%%%%%%%%%%%%%%%%%%%%%%%%
% References                                                       %
%%%%%%%%%%%%%%%%%%%%%%%%%%%%%%%%%%%%%%%%%%%%%%%%%%%%%%%%%%%%%%%%%%%%

\bibliography{KKbib.bib}

\end{document}